\documentclass[11pt]{amsart2000}
\usepackage{amsmath2000,graphicx}

\newtheorem{thm}{Theorem}[section]
\newtheorem{lem}[thm]{Lemma}
\newtheorem{cor}[thm]{Corollary}

\theoremstyle{definition}

\theoremstyle{remark}

\theoremstyle{plain}

\newcommand{\Z}{{\mathbb{Z}}}

\newcommand{\R}{{\mathbb{R}}}
\newcommand{\C}{{\mathbb{C}}}

\newcommand{\T}{{\mathbb{T}}}

\newcommand{\BB}{{\mathcal{B}}}

\numberwithin{equation}{section}

\begin{document}

\title[Norm estimates of almost Mathieu operators]
{Norm estimates of almost Mathieu operators}

\author[Boca and Zaharescu]{Florin P. Boca and Alexandru Zaharescu}

\address{Department of Mathematics, University of Illinois, Urbana, IL 61801}

\address{Institute of Mathematics of the Romanian Academy, P.O.Box 1-764,
Bucharest 70700, Romania}

\address{fboca@math.uiuc.edu \quad zaharesc@math.uiuc.edu}

\date{\today}
\begin{abstract}
We estimate the norm of the almost Mathieu operator 
$H_{\theta,\lambda} =U_\theta+U_\theta^*+\frac{\lambda}{2}
(V_\theta+V_\theta^*)$,
viewed as an element in the rotation $C^*$-algebra
$A_\theta =C^* (U_\theta ,V_\theta \ \mbox{\rm unitaries}\, ;\,
U_\theta V_\theta =e^{2\pi i\theta} V_\theta U_\theta)$. 
In the process, we prove the inequality
\begin{equation*}
\| H_{\theta,\lambda} \| \leq
\sqrt{4+\lambda^2-\bigg( 1-\frac{1}{\tan \pi \theta} \bigg)
\bigg( 1-\sqrt{\frac{1+\cos^2 4\pi \theta}{2}}\, \bigg)
\min (4,\lambda^2) } 
\end{equation*}
for every $\lambda \in \R$ and every $\theta \in [1/4,1/2]$.
This significantly improves the inequality $\| H_{\theta,2} \|
\leq 2\sqrt{2}$, $\theta \in [1/4,1/2]$, conjectured by
B\' eguin, Valette and Zuk.
\end{abstract}


\maketitle

\section{Introduction and Statement of Results}

An almost Mathieu operator is a discrete Schr\" odinger operator
that acts on the Hilbert space $\ell^2=\ell^2 (\Z)$ as
\begin{equation*}
H (\theta ,\lambda,\phi) \xi_n =\xi_{n+1}+\xi_{n-1}+\lambda\cos 
2\pi (n\theta +\phi) \, \xi_n ,
\end{equation*}
where $\theta$, $\lambda$ and $\phi$ are real numbers,
and $(\xi_n)_n$ 
denotes the canonical orthonormal basis in $\ell^2$. The study of the 
spectral properties of this class of operators has attracted a 
significant amount of interest in the past couple of decades 
(see \cite{BS}, \cite{Bel1}, \cite{CEY}, \cite{AMS}, \cite{HS}, \cite{L},
\cite{Zhi3}, \cite{BG} for some of the most important developments).
Most of this work has focused on the ``Ten Martini problem'' of
M. Kac, concerning the possible values of the labels of the
gaps that appear in the spectrum of these kinds of operators, and on
the localization properties of the spectrum.

The almost Mathieu operator $H(\theta,\lambda,\phi)$
 can be regarded as the image of the self-adjoint element
$H_{\theta,\lambda} =U+U^*+(\lambda /2)(V+V^*)$ in the 
representation of the rotation $C^*$-algebra 
$A_\theta =C^* (U,V$ unitaries ; $UV=e^{2\pi i\theta} VU)$ that maps 
$U$ to the bilateral shift $u_0$ defined on $\ell^2$ by 
$u_0 \xi_n =\xi_{n-1}$, and $V$ to the diagonal unitary $v_0$ defined 
by $v_0 \xi_n =e^{2\pi i (n\theta +\phi)} \xi_n$.
The operator $H_\theta =H_{\theta,2}$ is called a Harper operator.

\begin{figure}[ht]
\begin{center}
\includegraphics*[scale=0.8, bb=0 0 365 270]{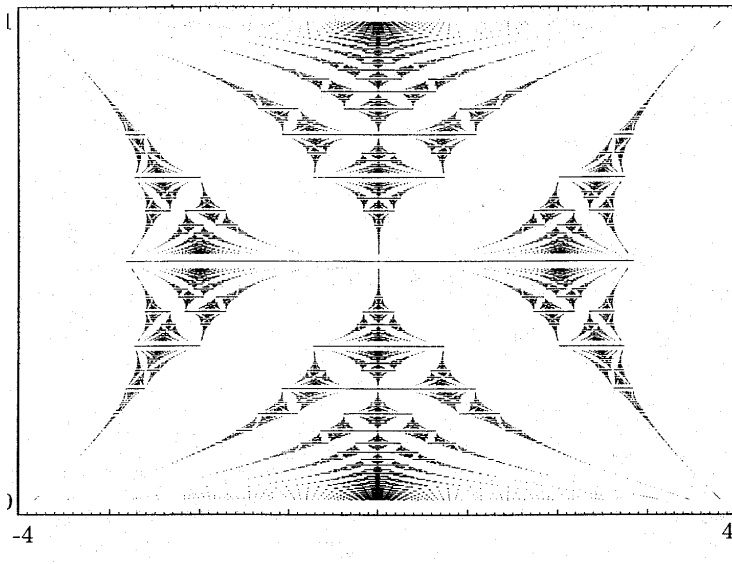}
\end{center}
\caption{The spectra of the Harper operators $H_{p/q}$
for $0\leq p/q \leq 1$} \label{Figure0}
\end{figure}

When $\theta =p/q$ is rational with $0\leq p<q$ coprime integers,
the spectrum of $H_{\theta,\lambda}$, viewed as an element of $A_\theta$,
consists either in the union of $q$ disjoint intervals (when $q$ is odd),
or of $q-1$ disjoint intervals (when $q$ is even). 
This is best illustrated in Hofstadter's butterfly (\cite{Hof}) in
Figure \ref{Figure0}.



The irreducible representations of $A_{p/q}$
can be classified up to unitary equivalence. They all have 
dimension $q$ and are given by
\begin{equation*}
\pi_{z_1,z_2} (U)=z_1 U_0 ,\qquad 
\pi_{z_1,z_2} (V)=z_2 V_0 
\end{equation*}
for $z_1 ,z_2 \in \T$, where
\begin{equation*}
\begin{split}
& U_0 =\left( \begin{matrix}
0 & 1 & 0 & \dots & 0 & 0 \\
0 & 0 & 1 & \dots & 0 & 0 \\
\vdots & \vdots & \vdots & \ddots & \vdots & \vdots \\
0 & 0 & 0 & \dots & 0 & 1 \\
1 & 0 & 0 & \dots & 0 & 0 
\end{matrix} \right) ,\quad
V_0 =\left( \begin{matrix} 
1 & & & & \\
 & e^{2\pi i\theta} & & & \\
 & & e^{4\pi i\theta} & & \\
 & &          & \ddots & \\
 & &          &  & e^{2\pi i(q-1)p/q} 
\end{matrix} \right) \\
& \mbox{\rm and} \ \theta =\frac{p}{q} \, .
\end{split} 
\end{equation*}
We also have in this case ($\theta$ rational) 
\begin{equation}\label{I1}
\| H_{\theta,\lambda} \| =\| h_{\theta,\lambda} \| ,
\end{equation}
where
\begin{equation}\label{I2}
h_{\theta,\lambda} =U_0+U_0^*+\frac{\lambda}{2} \, (V_0+V_0^*)
 \in M_q (\C) =\BB (\ell^2 (\Z_q)) .
\end{equation}

The map $\theta \mapsto \| H_{\theta,\lambda} \|$ is Lip$^{1/2}$ continuous
on $[0,1]$ (see \cite{AMS}, \cite{Bel2}, \cite{HR} or
\cite{KP} for different proofs with various degrees of generality).

All these properties of rotation $C^*$-algebras and of almost
Mathieu operators can be found, with self-contained proofs, in the 
first four chapters of \cite{Bo}.

The first estimates on the norm of almost Mathieu
operators were given by C. B\' eguin, A. Valette and
A. Zuk (\cite{BVZ}), who proved the inequality
\begin{equation}\label{I3}
\| H_\theta \| \leq 2(1+\sqrt{2}+\cos 2\pi \theta ) \qquad 
\theta \in [0,1].
\end{equation}
This is effective only in a neighbourhood of $\theta =1/2$, as
the norm of $H_\theta$ is no greater than $4$.

It was also conjectured by these authors in \cite{BVZ} that\footnote{
The existence of the $\ast$-antiautomorphism of $A_\theta$ that maps
$U$ to $U^*$ and $V$ to $V^*$ implies that the spectra of 
$H_{\theta,\lambda}$ and $H_{1-\theta,\lambda}$ coincide for any 
$\theta \in [0,1]$ and any $\lambda \in \R$.}
\begin{equation}\label{I4}
\| H_\theta \| \leq 2\sqrt{2} \qquad \theta \in [1/4,1/2 ].
\end{equation}

A proof of \eqref{I4} has appeared in \cite{Sz}. Unfortunately,
the trigonometric formulas at page 157 are not correct
and relation (13) in \cite{Sz} should change to
\begin{equation*}
(\sqrt{2}+s_n -\gamma_n )(\sqrt{2} -s_n +\gamma_{n-1}) \geq 1.
\end{equation*}
This makes the forthcoming 
arguments at pages 158-159, concerning the possibility of chosing 
$\gamma_n$ such that this inequality be satisfied, to be incorrect.

The original motivation of this work was to give a complete proof
of the inequality \eqref{I4}. This task is achieved first 
in Section 2, where we prove for every $\lambda \in \R$ and every 
$\theta \in [1/4,1/2]$ the inequality 
\begin{equation}\label{I5}
\| H_{\theta,\lambda} \| \leq \sqrt{4+\lambda^2} \, .
\end{equation}
Note that the equality holds in \eqref{I5} at $\theta =1/4$ and
$\theta =1/2$ for any $\lambda$. In this section we also prove 
in the range $\theta \in [0,1/2]$ the estimate
\begin{equation}\label{I6}
\| H_{\theta,\lambda} \| \leq M_\lambda (\theta)
=\sqrt{4+\lambda^2+4\vert \lambda \vert (\cos \pi \theta -\sin \pi 
\theta ) \cos \pi \theta} .
\end{equation}

\begin{figure}[ht]
\begin{center}
\includegraphics*[scale=1.5, bb=75 0 330 150]{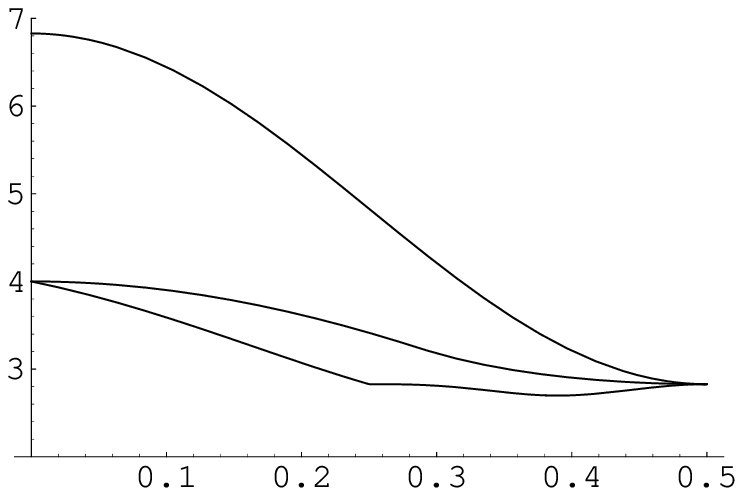}
\end{center}
\caption{The upper bound estimates \eqref{I3},\eqref{I8},\eqref{I6} and
\eqref{I7}}
\label{Figure2}
\end{figure}

\begin{figure}[ht]
\begin{center}
\includegraphics*[scale=1.5, bb=90 0 330 150]{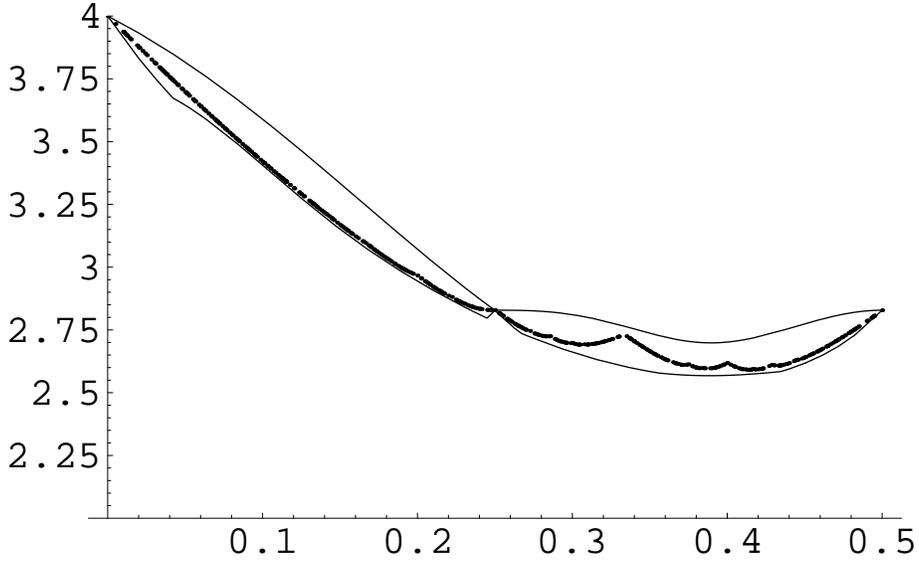}
\end{center}
\caption{The graphs of the functions $[0,1/2] \ni \theta
\mapsto M_2 (\theta),\| H_\theta \|,m(\theta)$}
\label{Figure1}
\end{figure}

In Section 3, we further sharpen the upper bound $\sqrt{4+\lambda^2}$
for $\| H_{\theta,\lambda} \|$ on $[1/4,1/2]$, and prove for
every $\theta \in [1/4,1/2]$ the inequality
\begin{equation}\label{I7}
\begin{split} 
& \| H_{\theta,\lambda} \| \leq M_\lambda (\theta) \\ & \qquad =
\sqrt{4+\lambda^2-\bigg( 1-\frac{1}{\tan \pi \theta} \bigg)
\bigg( 1-\sqrt{\frac{1+\cos^2 4\pi \theta}{2}}\, \bigg)
\min (4,\lambda^2) } ,
\end{split}
\end{equation}
which provides, when $\lambda =2$, a significant improvement of 
\eqref{I4} (see Figures \ref{Figure1} and \ref{Figure2}).
In the case $\lambda =2$, the upper bound estimates \eqref{I6} and 
\eqref{I7} are compared in Figure \ref{Figure2} with \eqref{I3} 
and with the bound
\begin{equation}\label{I8}
\| H_\theta \| \leq \begin{cases}
2+2\cos \pi \theta & \mbox{\rm if $\sin^2 \pi \theta \leq (\sqrt{5}-1)/2$}
\\ 2\sqrt{1+1/\sin^2 \pi \theta} & \mbox{\rm if $\sin^2 \pi \theta
\geq (\sqrt{5}-1)/2$,}
\end{cases}
\end{equation}  
proved (correctly) in \cite{Sz}.

In the last part of the paper, we give some explicit lower bounds
for the norm of Harper operators, proving the inequality
\begin{equation}\label{I9}
\| H_\theta \| \geq m(\theta)=\max \big(
f_1(\theta),f_2(\theta),f_3(\theta)\big) \qquad 
\theta \in [0,1/2 ],
\end{equation}
where $f_1 (\theta),f_2 (\theta),f_3(\theta) \geq 0$ are given by
\begin{equation}\label{I10}
\begin{split}
f_1(\theta)^2 & =6-\frac{4}{1+\frac{1}{\sin \pi \theta}+
\sqrt{1+\frac{1}{\sin \pi \theta}}}  \\
&  \qquad \qquad +\frac{2}{1+4\sin \pi \theta \cos^2 
\pi \theta}+\frac{8\cos^2 \pi \theta}{(1+\sin \pi \theta )^{3/2}} ,
\end{split}
\end{equation}

\begin{equation}\label{I11}
\begin{split}
&  \mbox{\small $\displaystyle
f_2(\theta)^2 =4+\frac{2}{\sqrt{1+\vert \sin 4\pi \theta \vert}}
+\frac{2\cos^2 2\pi \theta}{1+\vert \sin 4\pi \theta \vert}$}  \\
& \mbox{\small $\displaystyle 
+2\sqrt{\bigg( \frac{\sin^2 2\pi \theta}{1+\vert 
\sin 4\pi \theta \vert}\bigg)^2 
+\frac{16\cos^4 \pi \theta}{(2+\vert \tan 2\pi \theta \vert)^2}
\Big( 1+\frac{\cos 2\pi \theta}{\sqrt{1+\vert \sin 4\pi \theta \vert}}
\Big)^2}$} ,
\end{split}
\end{equation}

\begin{equation}\label{I12}
\begin{split}
& \mbox{\small $\displaystyle  f_3 (\theta)^2 =
4+\frac{4}{5}\, (\cos 2\pi \theta +2\cos^2 2\pi \theta 
+2\cos^4 2\pi \theta )$} \\ & 
\mbox{\small $\displaystyle +\sqrt{\bigg( 2-
\frac{4(\cos 2\pi \theta +2\cos^2 2\pi \theta 
+2\cos^4 2\pi \theta )}{5} \bigg)^2 +
\bigg( \sqrt{10}+\frac{8\cos 2\pi \theta}{\sqrt{10}} \bigg)^2}.$}
\end{split}
\end{equation}

In particular we get
\begin{equation}\label{I13}
\min\limits_{1/4\leq \theta \leq 1/2}
 \| H_\theta \| \geq \min\limits_{1/4<\theta <1/2}
f_1 (\theta ) \approx \sqrt{6.59303} \approx 2.56769.
\end{equation}
This is a good estimate for $\min\limits_{1/4\leq \theta \leq 1/2}
 \| H_\theta \|$, which appears to be, by numerical computations, 
just fractionally larger than $2.59$.

We also get (see Figure \ref{Figure1})
\begin{equation}\label{I14}
\min\limits_{0\leq \theta \leq 1/4} \| H_\theta \|^2 \geq 7.82387
\end{equation}
and for all $0\leq \theta \leq 0.23441$ we get
\begin{equation}\label{I15}
\| H_\theta \|^2 \geq 8 .
\end{equation}

Our estimates \eqref{I6}, \eqref{I7} and \eqref{I9} on the norm of
the Harper operator $H_\theta$ are illustrated in Figure \ref{Figure1}. 

Numerical computations appear to indicate that inequality \eqref{I15} 
might hold for every $\theta \in [0,1/4]$. It would be interesting 
to clarify this point.

\section{A proof of the inequality $\| H_{\theta,\lambda} \| \leq 
\sqrt{4+\lambda^2}$, 
$1/4 \leq \theta \leq 1/2$}

Throughout this section, we take $\theta =p/q$, denote
\begin{equation*}
C_n =C_n (\theta)=\cos 2n \pi \theta \qquad n\in \Z ,
\end{equation*}
and consider the self-adjoint $q\times q$ matrix $h_{\theta,\lambda}$ 
defined in \eqref{I2}. That is, $h_{\theta,\lambda}$ acts on 
$\ell^2 (\Z_q)$ by
\begin{equation*}
h_{\theta,\lambda} \epsilon_n =\epsilon_{n+1} +\epsilon_{n-1}
+\lambda C_n \epsilon_n \qquad n\in \Z_q ,
\end{equation*}
where $( \epsilon_n )_{n\in \Z_q}$ denotes the canonical 
orthonormal basis in $\ell^2 (\Z_q)$.

In this section and the next one, we set 
$\sum\limits_m =\sum\limits_{m\in \Z_q}$.

Let $(X_m)_{m\in \Z_q}$ be a unit eigenvector in $\ell^2 (\Z_q)$
for the eigenvalue 
$E$ of $h_{\theta,\lambda}$ with $X_m=X_m (\theta) \in \R$. Then
\begin{equation}\label{2.1}
X_{m+1}+X_{m-1} +\lambda C_m X_m =E X_m  \qquad m\in \Z_q ,
\end{equation}
which gives
\begin{equation}\label{2.2}
\begin{split}
E^2 & =\sum\limits_m (E X_m)^2 =\sum\limits_m 
(X_{m+1}+X_{m-1}+\lambda C_m X_m )^2 \\
& =2+\lambda^2 \sum\limits_m C_m^2 X_m^2 +
2\sum\limits_m X_{m+1} X_{m-1} \\ & \qquad +2\lambda \sum\limits_m
X_m X_{m-1} (C_m+C_{m-1}) \\
& =2+\lambda^2 \sum\limits_m (1-\sin^2 2m \pi \theta )X_m^2 
+2-\sum\limits_m (X_{m+1}-X_{m-1})^2 \\ & \qquad  +2\lambda \sum\limits_m
X_m X_{m-1} (C_m+C_{m-1}) \\
& =4+\lambda^2 -\sum\limits_m (X_{m+1}-X_{m-1})^2 -
\lambda^2 \sum\limits_m X_m^2 \sin^2 2m \pi \theta \\ & \qquad +
2\lambda \sum\limits_m X_m X_{m-1} (C_m+C_{m-1}).
\end{split}
\end{equation}

It also follows from \eqref{2.1} that
\begin{equation}\label{2.3}
\begin{split}
\lambda \sum\limits_m & X_m X_{m-1} C_{m-1} =\sum\limits_m X_m (E 
X_{m-1} -X_m -X_{m-2}) \\
& =E \sum\limits_m X_m X_{m-1} -1-\sum\limits_m X_m X_{m-2} \\
& =E \sum\limits_m X_m X_{m-1} -\sum\limits_m X_{m-1}^2 -
\sum\limits_m X_{m+1} X_{m-1} \\ 
& =\sum\limits_m X_{m-1} (E X_m -X_{m-1}-X_{m+1}) \\
& =\lambda \sum\limits_m X_m X_{m-1} C_m .
\end{split}
\end{equation}

Subtracting the right-hand side from the left-hand side 
in \eqref{2.3}, we get
\begin{equation}\label{2.4}
\sum\limits_m X_m X_{m-1} \sin (2m-1)\pi \theta =0 .
\end{equation}

By \eqref{2.2} we infer that
\begin{equation}\label{2.5}
\begin{split}
E^2 & = 4+\lambda^2 -\sum\limits_m (X_{m+1}-X_{m-1}
+\lambda X_m \sin 2m\pi \theta)^2
\\ & \quad +2\lambda \sum\limits_m X_m X_{m-1} (C_m+C_{m-1}) 
+2\lambda \sum\limits_m X_{m+1} X_m \sin 2m\pi \theta \\ & \qquad -
2\lambda \sum\limits_m X_m X_{m-1} \sin 2m\pi \theta \\
& =4+\lambda^2 -\sum\limits_m (X_{m+1}-X_{m-1}+\lambda 
X_m \sin 2m\pi \theta)^2 
\\ & \quad +2\lambda \sum\limits_m X_m X_{m-1} \big( C_m+C_{m-1}
+\sin 2(m-1)\pi \theta -\sin 2m\pi \theta \big).
\end{split}
\end{equation}

But
\begin{equation*}
\begin{split}
& \cos 2m\pi \theta +\cos 2(m-1)\pi \theta +\sin 2(m-1)\pi \theta -
\sin 2m\pi \theta \\
& \quad = \cos 2m\pi \theta +\cos 2m\pi \theta \cos 2\pi \theta +
\sin 2m\pi \theta \sin 2\pi \theta \\
& \qquad +\sin 2m\pi \theta \cos 2\pi \theta 
-\cos 2m\pi \theta \sin 2\pi \theta -\sin 2m\pi \theta \\
& \quad =\cos 2m\pi \theta (1+\cos 2\pi \theta -\sin 2\pi \theta )
-\sin 2m\pi \theta (1-\cos 2\pi \theta -\sin 2\pi \theta ) \\
& \quad =2\cos 2m\pi \theta \cos \pi \theta (\cos \pi \theta -\sin 
\pi \theta ) -2\sin 2m\pi \theta \sin \pi \theta (\sin \pi \theta 
-\cos \pi \theta ) \\
& \quad =2(\cos \pi \theta -\sin \pi \theta ) 
(\cos \pi \theta \cos 2m\pi \theta +\sin \pi \theta \sin 2m\pi \theta ) \\
& \quad =2(\cos \pi \theta -\sin \pi \theta ) \cos (2m-1) \pi \theta ,
\end{split}
\end{equation*}
which gives in conjunction with \eqref{2.5}
\begin{equation}\label{2.6}
\begin{split}
E^2 & =4+\lambda^2 -\sum\limits_m (X_{m+1}-X_{m-1}
+\lambda X_m \sin 2m\pi \theta )^2 \\ & 
\qquad \qquad +4\lambda (\cos \pi \theta -\sin \pi \theta )
\sum\limits_m X_m X_{m-1} \cos (2m-1) \pi \theta .
\end{split}
\end{equation}

Using also
\begin{equation*}
C_m+C_{m-1} =\cos 2m\pi \theta +\cos 2(m-1)\pi \theta =2\cos \pi \theta
\cos (2m-1)\pi \theta ,
\end{equation*}
we derive from \eqref{2.6} and \eqref{2.3}
\begin{equation}\label{2.7}
\begin{split}
E^2 & =4+\lambda^2 -\sum\limits_m (X_{m+1}-X_{m-1} 
+\lambda X_m \sin 2m\pi \theta)^2
\\ & \qquad +4\lambda (1-\tan \pi \theta )\sum\limits_m X_m X_{m-1} C_m 
\qquad 0\leq \theta <1/2 .
\end{split}
\end{equation}

From \eqref{2.6} it follows that
\begin{equation}\label{2.8}
E^2 \leq 4+\lambda^2+4\lambda (\cos \pi \theta -\sin \pi \theta ) 
\sum\limits_m X_m X_{m-1} \cos (2m-1) \pi \theta .
\end{equation}

Using the identity
\begin{equation*}
2(ax+by)=(a+b)(x+y)+(a-b)(x-y)
\end{equation*}
for $a=C_{m+1}$, $b=C_{m-1}$, $x=X_{m+1}$,
$y=X_{m-1}$, we may write
\begin{equation}\label{2.9}
\begin{split}
2\lambda & \sum\limits_m X_m X_{m-1} (C_m+C_{m-1}) \\ & =
\lambda \sum\limits_m X_m X_{m-1}(C_m+C_{m-1}) +\lambda \sum\limits_m
X_{m+1} X_m (C_{m+1} +C_m) \\
& =\lambda \sum\limits_m X_m (X_{m+1}+X_{m-1}) C_m \\ & \qquad +
\frac{\lambda}{2} \sum\limits_m X_m (X_{m+1}+X_{m-1})(C_{m+1}+C_{m-1}) \\
& \qquad +\frac{\lambda}{2}
\sum\limits_m X_m (X_{m+1}-X_{m-1}) (C_{m+1} -C_{m-1})
\\ & =\frac{\lambda}{2}
\sum\limits_m X_m (X_{m+1}+X_{m-1})(C_{m+1}+2C_m +C_{m-1}) \\
& \qquad +\frac{\lambda}{2}
\sum\limits_m X_m (X_{m+1}-X_{m-1})(C_{m+1}-C_{m-1}).
\end{split}
\end{equation}

Since
\begin{equation*}
C_{m+1}-C_{m-1} =\cos 2(m+1)\pi \theta -\cos 2(m-1)\pi \theta =
-2\sin 2\pi \theta \sin 2m\pi \theta ,
\end{equation*}
we collect from \eqref{2.2} and \eqref{2.9}
\begin{equation}\label{2.10}
\begin{split}
E^2 & =4+\lambda^2+\frac{\lambda}{2}
\sum\limits_m X_m (X_{m+1}+X_{m-1}) (C_{m+1}+2C_m +C_{m-1})
\\ & \qquad -\lambda^2 \sum\limits_m X_m^2 \sin^2 2m\pi \theta 
-\sum\limits_m (X_{m+1}-X_{m-1})^2 \\ & \qquad -\lambda \sin 2\pi \theta 
\sum\limits_m X_m (X_{m+1}-X_{m-1})\sin 2m\pi \theta . 
\end{split}
\end{equation}
Using
\begin{equation*}
\begin{split}
& \sum\limits_m X_m (X_{m+1}+X_{m-1}) (C_{m+1}+2C_m +C_{m-1}) \\
& =\sum\limits_m \Big( X_{m-1} X_m (C_m +2C_{m-1}+C_{m-2}) +
 X_{m-1} X_m (C_{m+1}+2C_m +C_{m-1})\Big)  \\
& =\sum\limits_m X_m X_{m-1} (C_{m+1}+3C_m +3C_{m-1}+C_{m-2}),
\end{split}
\end{equation*}
\begin{equation*}
\begin{split}
 C_{m+1} +3C_m +3 & C_{m-1}+C_{m-2} \\ &  =\Re ( 
e^{2\pi i(m+1)\theta} +3e^{2\pi i m\theta}
+3e^{2\pi i(m-1)\theta} +e^{2\pi i(m-2)\theta} ) \\ & 
= \Re \big( e^{2\pi i(m-2)\theta} ( 
1+e^{2\pi i\theta} )^3 \big) =\Re \big( e^{(2m-1)\pi i\theta} 
(e^{-\pi i\theta} +e^{\pi i\theta} )^3 \big) 
\\ & =\Re \big( e^{(2m-1)\pi i\theta} 
(2\cos \pi \theta )^3 \big) =8\cos^3 \pi \theta \cos (2m-1)\pi \theta 
\end{split} 
\end{equation*}
and \eqref{2.10}, we collect
\begin{equation}\label{2.11}
\begin{split}
E^2 & =4+\lambda^2+4\lambda \cos^3 \pi \theta \sum\limits_m X_m X_{m-1}
\cos (2m-1)\pi \theta \\ & \qquad -\lambda^2
\sum\limits_m X_m^2 \sin^2 2m\pi \theta 
-\sum\limits_m (X_{m+1}-X_{m-1})^2 \\ & \qquad  
-\lambda \sin 2\pi \theta \sum\limits_m X_m (X_{m+1}-X_{m-1}) 
\sin 2m\pi \theta \\
& =4+\lambda^2+4\lambda \cos^3 \pi \theta \sum\limits_m X_m X_{m-1} 
\cos (2m-1)\pi \theta  \\ & \qquad
-\frac{\lambda^2}{4} \,
(4-\sin^2 2\pi \theta ) \sum\limits_m X_m^2 \sin^2 2m\pi \theta
\\ & \qquad -\sum\limits_m \bigg( \frac{\lambda}{2}\, 
X_m \sin 2\pi \theta \sin 2m\pi \theta  +X_{m+1}-X_{m-1}\bigg)^2 \\  
& \leq 4+\lambda^2 +4\lambda \cos^3 \pi \theta \sum\limits_m X_m X_{m-1}
\cos (2m-1)\pi \theta .
\end{split}
\end{equation}

We can now prove

\medskip

\begin{thm}\label{T2.1}
{\em (i)} For every $\theta \in [1/4,1/2]$, we have
\begin{equation*}
\| H_{\theta,\lambda} \| \leq \sqrt{4+\lambda^2} .
\end{equation*}

{\em (ii)} For every $\theta \in [0,1/4]$, we have
\begin{equation*}
\| H_{\theta,\lambda} \|^2 \leq 4+\lambda^2 +
4\lambda (\cos \pi \theta -\sin \pi \theta)\cos \pi \theta.
\end{equation*}
\end{thm}

{\sl Proof.}
Since $H_{\theta,\lambda}$ and $H_{\theta,-\lambda}$ have the 
same spectrum, we may assume that $\lambda >0$.
Using the continuity of the map $\theta \mapsto \| H_{\theta,\lambda} \|$,
it suffices to assume in the first place that $\theta$ is rational.
Taking now stock on \eqref{I1}, we may replace
$\| H_{\theta,\lambda} \|$ by $\| h_{\theta,\lambda} \|$.

(i) Let $\theta \in [1/4,1/2)$. The sum
$\sum_m X_m X_{m-1} \cos (2m-1)\pi \theta$ played a central role in the 
previous computations. We do not have any control over this sum.
However, the point is that if $\sum_m X_m X_{m-1}
\cos (2m-1)\pi \theta \leq 0$, then \eqref{2.11} gives $E^2 \leq 
4+\lambda^2$, and if $\sum_m X_m X_{m-1} \cos (2m-1)\pi 
\theta \geq 0$, then \eqref{2.8} gives $E^2 \leq 4+\lambda^2 $.

(ii) This estimate follows from \eqref{2.8} and from the next lemma.
\qed

\medskip

\begin{lem}\label{L2.2}
For every $\theta \in [0,1/2]$, we have
\begin{equation*}
\begin{split}
\left| \sum\limits_m X_m X_{m-1} \cos (2m-1)\pi \theta \right| & \leq 
\frac{1}{2} \sqrt{2(1+\vert \cos 2\pi \theta \vert)} \\ & =
\begin{cases} \cos \pi \theta & \mbox{if $0\leq \theta \leq 1/4$} \\
\sin \pi \theta & \mbox{if $1/4 \leq \theta \leq 1/2$.}
\end{cases}
\end{split}
\end{equation*}
\end{lem}

{\sl Proof.}
With $D_m =\cos (2m-1)\pi \theta$, we gather
\begin{equation*}
\begin{split}
& D_m^2+D_{m+1}^2 =\cos^2 (2m-1)\pi \theta +\cos^2 (2m+1)\pi \theta 
\\ & \quad =1+\big( \cos (4m-2)\pi \theta +\cos (4m+2) \pi \theta \big)/2
 =1+\cos 4m\pi \theta \cos 2\pi \theta \\
& \leq 1+\vert \cos 2\pi \theta \vert .
\end{split}
\end{equation*}
Combining this with Cauchy-Schwartz, we derive
\begin{equation*}
\begin{split}
2\sum\limits_m \vert X_m X_{m-1} D_m \vert & =
\sum\limits_m \big( \vert X_m \vert \vert X_{m-1} \vert
\vert D_m \vert +\vert X_{m+1} \vert \vert X_m \vert \vert D_{m+1} \vert 
\big) \\
& =\sum\limits_m \vert X_m \vert \big( \vert X_{m-1} \vert \vert D_m \vert 
+\vert X_{m+1} \vert \vert D_{m+1} \vert \big) \\
& \leq \sum\limits_m \vert X_m \vert \sqrt{X_{m-1}^2+X_{m+1}^2}\
\sqrt{D_m^2+D_{m+1}^2} \\
& \leq \sqrt{1+\vert \cos 2\pi \theta \vert} \ \sum\limits_m
\vert X_m \vert \ \sqrt{X_{m-1}^2 +X_{m+1}^2} \\
& \leq \sqrt{1+\vert \cos 2\pi \theta \vert}\ 
\sqrt{\sum\limits_m X_m^2} \
\sqrt{ \sum\limits_m (X_{m-1}^2+X_{m+1}^2 )} \\
& =\sqrt{2(1+\vert \cos 2\pi \theta \vert )} .
\end{split}
\end{equation*}
\qed

\bigskip

\section{Improved upper bound estimates in the range
$\theta \in [1/4,1/2]$}

Since the spectra of $H_{\theta,\lambda}$ and $H_{\theta,-\lambda}$
coincide, we may assume that $\lambda > 0$. We also assume throughout 
this section that $\theta \in (1/4,1/2)$.

Using the notation from the previous section, we set
\begin{equation*}
\begin{split}
& S=S(\theta,\lambda,E)=E \sum\limits_m C_m X_m^2
+\frac{2}{\lambda} \sum\limits_m X_{m+1} X_{m-1} ,\\
& T=T(\theta,\lambda)=\lambda \sum\limits_m C_m^2 X_m^2
+\frac{2}{\lambda} \sum\limits_m X_{m+1} X_{m-1}  .
\end{split}
\end{equation*}

It is worth to note first that
\begin{equation*}
\begin{split}
\sum\limits_m X_m X_{m-1} & =\sum\limits_m X_{m+1} X_m
=\sum\limits_m \big( (E -\lambda C_m )X_m -X_{m-1} \big) X_m \\
& =E-\lambda \sum\limits_m C_m X_m^2 -\sum\limits_m X_m X_{m-1} ,
\end{split}
\end{equation*}
thus
\begin{equation}\label{3.1}
\sum\limits_m X_m X_{m-1} =\sum\limits_m X_{m+1} X_m =
\frac{E}{2} -\frac{\lambda}{2} \sum\limits_m C_m X_m^2 ,
\end{equation}
and as a result
\begin{equation}\label{3.2}
\begin{split}
2\lambda \sum\limits_m X_m X_{m-1} & C_m  
=\lambda \sum\limits_m (X_{m+1} X_m C_{m+1} +X_m X_{m-1} C_m) \\
\mbox{\rm by \eqref{2.3}} \quad 
& =\lambda \sum\limits_m X_{m+1} X_m C_m 
+\lambda \sum\limits_m X_m X_{m-1} C_m \\
& =\lambda \sum\limits_m C_m X_m (X_{m+1}+X_{m-1}) \\
\mbox{\rm by \eqref{2.1}} \quad & =\sum\limits_m 
(E X_m -X_{m+1}-X_{m-1})(X_{m+1}+X_{m-1}) \\
& =2E \sum\limits_m X_{m+1} X_m -\sum\limits_m 
(X_{m+1} +X_{m-1})^2 \\
\mbox{\rm by \eqref{3.1}} \quad & =2E \bigg(
\frac{E}{2} -\frac{\lambda}{2} \sum\limits_m C_m X_m^2 \bigg) 
-2-2\sum\limits_m X_{m+1} X_{m-1} \\
& =E^2 -\lambda E \sum\limits_m C_m X_m^2 -
2\sum\limits_m X_{m+1} X_{m-1} -2 \\
& =E^2 -\lambda S -2 .
\end{split}
\end{equation}

Therefore, \eqref{2.7} implies 
\begin{equation*}
E^2 \leq 4+\lambda^2 +2(1-\tan \pi \theta)(E^2 -\lambda S -2) ,
\end{equation*}
and so,
\begin{equation*}
E^2 \leq \frac{\lambda^2 +2\lambda (\tan \pi \theta -1)S
+4\tan \pi \theta}{2\tan \pi \theta -1} \, ,
\end{equation*}
which gives in turn
\begin{equation}\label{3.3}
E^2 -4-\lambda^2 \leq \frac{2(\tan \pi \theta -1)
(\lambda S-\lambda^2 -2)}{2\tan \pi \theta -1} \, .
\end{equation}

\medskip

\begin{lem}\label{L3.1}
Suppose that $(Y_m)_{m\in \Z_q}$ is a unit vector in $\ell^2 (\Z_q)$. Then
\begin{equation*}
2\sum\limits_m C_m^2 Y_m^2 +\sum\limits_m Y_{m+1} Y_{m-1} 
\leq 1+\sqrt{2(1+\cos^2 4\pi \theta )} .
\end{equation*}
\end{lem}

{\sl Proof.}
If we write
\begin{equation*}
2\sum\limits_m C_m^2 Y_m^2 =1+\sum\limits_m Y_m^2 \cos 4m\pi \theta ,
\end{equation*}
then the desired result will follow at once as we have shown that
\begin{equation}\label{3.4}
\sup\limits_m \frac{Y_{m+1}^2 \cos 4(m+1)\pi \theta +
Y_{m-1}^2 \cos 4(m-1)\theta +2Y_{m+1}Y_{m-1}}{Y_{m+1}^2+Y_{m-1}^2} 
\end{equation}
is less or equal than $\sqrt{2(1+\cos^2 4\pi \theta)}$.
Taking $A=Y_{m+1} /\sqrt{Y_{m+1}^2+Y_{m-1}^2}$ and
$B=Y_{m-1}/\sqrt{Y_{m+1}^2+Y_{m-1}^2}$, the expression in \eqref{3.4}
is no greater than
\begin{equation*}
\begin{split}
& \sup\limits_{\substack{m\in \Z \\ A^2+B^2=1}} \hspace{-5pt}  
\Big( (A^2+B^2)\cos 4\pi \theta \cos 4m\pi \theta 
+(B^2-A^2) \sin 4\pi \theta \sin 4m\pi \theta +2AB \Big) \\ 
& \qquad = \sup\limits_{A^2+B^2=1} 
\left( \sqrt{(A^2+B^2)^2 \cos^2 4\pi \theta 
+(B^2-A^2)^2 \sin^2 4\pi \theta }+2AB \right) 
\\ & \qquad \qquad =\sup\limits_{A^2+B^2=1} \left( 
\sqrt{1-4A^2 B^2 \sin^2 4\pi \theta} +2AB \right) .
\end{split}
\end{equation*}
The statement follows now from 
\begin{equation*}
\begin{split}
\big( & \sqrt{1-4A^2B^2 \sin^2 4\pi \theta}+2AB \big)^2 \leq
2(1-4A^2B^2 \sin^2 4\pi \theta +4A^2 B^2) \\ & \qquad \qquad =
2(1+4A^2 B^2 \cos^2 4\pi \theta ) \leq 2(1+\cos^2 4\pi \theta ) .
\end{split}
\end{equation*}
\qed

\medskip

\begin{cor}\label{C3.2}
If $(Y_m)_{m\in \Z_q}$ is a unit vector in
$\ell^2 (\Z_q)$ and $0\leq \lambda \leq 2$, then
\begin{equation*}
\lambda \sum\limits_m C_m^2 Y_m^2 +
\frac{2}{\lambda} \sum\limits_m Y_{m+1} Y_{m-1}
\leq \frac{2}{\lambda} +\lambda 
\sqrt{\frac{1+\cos^2 4\pi \theta}{2}} \, .
\end{equation*}
\end{cor}

{\sl Proof.}
Using lemma \ref{L3.1} and Cauchy-Schwarz, the expression in the 
left-hand side becomes
\begin{equation*}
\begin{split}
& \frac{\lambda}{2} \bigg( 2\sum\limits_m C_m^2 Y_m^2
+\sum\limits_m Y_{m+1} Y_{m-1} \bigg) +
\bigg( \frac{2}{\lambda}-\frac{\lambda}{2} \bigg) 
\sum\limits_m Y_{m+1} Y_{m-1} \\
& \quad \leq \frac{2}{\lambda}-\frac{\lambda}{2} +
\frac{\lambda}{2} \, \big( 1+\sqrt{2(1+\cos^2 4\pi \theta )}\,  \big) 
=\frac{2}{\lambda} +\lambda \sqrt{\frac{1+\cos^2 4\pi \theta}{2}} 
\, .
\end{split}
\end{equation*}
\qed

\medskip

\begin{thm}\label{T3.3}
For every $\theta \in [1/4,1/2]$ and $\lambda \in \R$, we have 
\begin{equation*}
\| H_{\theta,\lambda} \| \leq 
\sqrt{4+\lambda^2-\bigg( 1-\frac{1}{\tan \pi \theta}
\bigg) \bigg( 1-\sqrt{\frac{1+\cos^2 4\pi \theta}{2}}\  \bigg)
\min (4,\lambda^2)} \, .
\end{equation*}
\end{thm}

{\sl Proof.}
We may assume that $\lambda > 0$.
We may also assume as in the proof of Theorem \ref{T2.1} that
$\theta$ is rational and then replace $\| H_{\theta,\lambda} \|$ by 
$\| h_{\theta,\lambda} \|$. Let $\varepsilon$ be a constant to be 
chosen later. 

We first consider the case $\lambda \leq 2$. Suppose that
\begin{equation}\label{3.5}
2\lambda \sum\limits_m X_m X_{m-1} C_m \leq \varepsilon .
\end{equation}

Since
\begin{equation*}
\begin{split}
2\sum\limits_m X_m X_{m-1} C_m & =\sum\limits_m X_m X_{m-1} C_m +
\sum\limits_m X_{m+1} X_m C_m \\
& =\sum\limits_m C_m X_m (X_{m-1}+X_{m+1}) =
\sum\limits_m (E-\lambda C_m) C_m X_m^2 \\
& =E\sum\limits_m C_m X_m^2 -\lambda \sum\limits_m C_m^2 X_m^2 \\
& =S-T ,
\end{split}
\end{equation*}
we infer from \eqref{3.5} that
\begin{equation}\label{3.6}
S\leq \frac{\varepsilon}{\lambda}+T .
\end{equation}

We combine \eqref{3.6} with \eqref{3.3} to collect
\begin{equation}\label{3.7}
E^2 -4-\lambda^2 \leq \frac{2(\tan \pi \theta -1)}{2\tan \pi \theta -1}
\, (\varepsilon +\lambda T -\lambda^2 -2) .
\end{equation}

If
\begin{equation*}
2\lambda \sum\limits_m X_m X_{m-1} C_m \geq \varepsilon ,
\end{equation*}
then we infer from \eqref{2.7} and from
$1-\tan \pi \theta \leq 0$ that
\begin{equation}\label{3.8}
E^2 -4-\lambda^2 \leq 4\lambda (1-\tan \pi \theta)
\sum\limits_m X_m X_{m-1} C_m \leq -2(\tan \pi \theta -1) \varepsilon .
\end{equation}

In summary, we combine \eqref{3.7} with \eqref{3.8} to get
\begin{equation*}
E^2 -4-\lambda^2 \leq 2(\tan \pi \theta -1) 
\max \bigg( -\varepsilon ,\frac{\varepsilon +\lambda T-\lambda^2 -2}{
2\tan \pi \theta -1} \bigg)
\end{equation*}
for any $\varepsilon \in \R$. The inequality
\begin{equation}\label{3.9}
\| H_{\theta ,\lambda} \| \leq \sqrt{4+\lambda^2 -
\bigg( 1-\frac{1}{\tan \pi \theta} \bigg) \bigg( 1-
\sqrt{\frac{1+\cos^2 4\pi \theta}{2}} \, \bigg) \lambda^2}
\end{equation}
follows by chosing
\begin{equation*}
\varepsilon =\frac{\lambda^2 +2-\lambda T}{2\tan \pi \theta} 
\end{equation*}
and employing Corollary \ref{C3.2}.

When $\lambda >2$, the desired inequality is obtained combining
\eqref{3.9} with the Andre-Aubry duality
\begin{equation*}
\| H_{\theta ,\lambda} \| =\frac{\lambda}{2} \, \| 
H_{\theta ,\frac{4}{\lambda}} \| .
\end{equation*}
\qed

\bigskip

\section{Some lower bounds for the norm of a Harper operator}

We shall consider sequences $(x_n)_n$ and 
$(y_n)_n$  in $\ell^2$, of the form
\begin{equation*}
x_n =x_n (\theta)=r^{\vert n\vert}
\end{equation*}
and
\begin{equation*}
y_n =y_n (\theta )=\begin{cases}
Ar^{\vert k\vert} , & \mbox{\rm if $n=2k$} \\
Br^{\vert k\vert} , & \mbox{\rm if $n=2k+1$} 
\end{cases} ,
\end{equation*}
with $A^2+B^2=1$, and with $0<r=r(\theta)<1$ to be chosen later. We have
\begin{equation*}
\sum\limits_n x_n^2 =\sum\limits_n y_n^2 =
\sum\limits_n r^{2\vert n\vert} =\frac{1+r^2}{1-r^2} \, ,
\end{equation*}
where we set $\sum\limits_n =\sum\limits_{n\in \Z}$.

Using the relations
\begin{equation}\label{4.1}
\sum\limits_k r^{\vert k\vert} \cos (ak+b)
=\frac{(1-r^2)\cos b}{1-2r\cos a+r^2} \, ,
\end{equation}
\begin{equation}\label{4.2}
\sum\limits_k r^{\vert k\vert+\vert k-1 \vert} \cos (ak+b)=
\frac{2r(1-r^2)\cos \frac{a}{2} \cos \big( b+\frac{a}{2} \big)}{1
-2r^2 \cos a+r^4} \, ,
\end{equation}
and
\begin{equation*}
\sum\limits_k r^{\vert k+1\vert+\vert k-1\vert} =
r^2 -1+\sum\limits_k r^{2\vert k\vert},
\end{equation*}
we find
\begin{equation*}
\begin{split}
& \| H_\theta \|^2  \geq \frac{1-r^2}{1+r^2} \sum\limits_n 
(2x_n \cos 2n\pi \theta +x_{n+1}+x_{n-1})^2 \\
& =\frac{1-r^2}{1+r^2} \sum\limits_n \big( 2r^{\vert n\vert}
\cos 2n\pi \theta +r^{\vert n+1\vert} +r^{\vert n-1 \vert} \big)^2 \\
& =6-\frac{(1-r^2)^2}{1+r^2} +\frac{2(1-r^2)^2}{1-2r^2 \cos 4\pi \theta
+r^4} +8\, \frac{2r}{1+r^2} \cdot 
\frac{(1-r^2)^2 \cos^2 \pi \theta}{1-2r^2 \cos 2\pi \theta +r^4} \, .
\end{split}
\end{equation*}

Taking $r=\tan (\alpha /2)$ with $0<\alpha<\pi /2$, it follows that
$\| H_\theta \|^2$ is greater or equal than
\begin{equation*}
\sup\limits_{0<\alpha <\pi/2}
\Bigg( 6-\frac{4\cos^2 \alpha}{1+\cos \alpha} +2\cos^2 \alpha
\bigg( \frac{1}{1-\sin^2 \alpha \cos^2 2\pi \theta}
+\frac{4\sin \alpha \cos^2 \pi \theta}{1-\sin^2 \alpha \cos^2 \pi \theta}
\bigg) \Bigg) .
\end{equation*}

Chosing $\alpha$ such that
\begin{equation*}
\cos^2 \alpha =\frac{\sin \pi \theta}{1+\sin \pi \theta} \, ,
\end{equation*}
we arrive at
\begin{equation}\label{4.3} 
\| H_\theta \|^2 \geq f_1 (\theta )^2 \qquad 
\theta \in [0,1/2] ,
\end{equation}
where $f_1 (\theta)$ is defined in \eqref{I10}.

In particular, we infer from \eqref{4.3} and \eqref{I10} that
\begin{equation}\label{4.4}
\min\limits_{1/4 \leq \theta \leq 1/2} \| H_\theta \| \geq 
\min\limits_{1/4< \theta < 1/2} f_1 (\theta) \approx
\sqrt{6.59303} \approx 2.56769 .
\end{equation}

To get a better estimate in a neighborhood of $1/4$ and $1/2$, we use the 
sequence $(y_n)_n$ and relations \eqref{4.1} and \eqref{4.2}, to collect
\begin{equation*}
\begin{split}
\sum\limits_n & (2y_n \cos 2n\pi \theta +y_{n+1}+y_{n-1} )^2 =
\sum\limits_k \Big( 2Ar^{\vert k\vert} \cos 4k\pi \theta +B
(r^{\vert k\vert}+r^{\vert k-1} )\Big)^2 \\
& \qquad \qquad +\sum\limits_k \Big( 2Br^{\vert k\vert} 
\cos (4k+2)\pi \theta +A (r^{\vert k\vert} +r^{\vert k+1\vert} )\Big)^2 \\
& =4A^2 \sum\limits_k r^{2\vert k\vert} \cos^2 4k\pi \theta +
4B^2 \sum\limits_k \cos^2 (4k+2)\pi \theta \\ & \qquad +2(A^2+B^2)
\sum\limits_k (r^{2\vert k\vert}+r^{\vert k\vert +\vert k-1\vert}) \\
& \qquad +4AB \sum\limits_k r^{2\vert k\vert}
\big(\cos 4k\pi \theta+\cos (4k+2)\pi \theta \big) \\ & \qquad
+4AB \sum\limits_k r^{\vert k\vert+\vert k-1\vert}
\big( \cos 4k\pi \theta +\cos (4k-2)\pi \theta \big) .   
\end{split}
\end{equation*}
This further equals
\begin{equation*}
\begin{split}
4\sum\limits_k & r^{2\vert k\vert} +2A^2 \sum\limits_k 
r^{2\vert k\vert} \cos 8k\pi \theta +2B^2 \sum\limits_k
r^{2\vert k\vert} \cos (8k+4)\pi \theta \\ & \qquad +2(A^2+B^2)
\sum\limits_k r^{\vert k\vert+\vert k-1\vert} 
+8AB \cos \pi \theta \sum\limits_k 
r^{2\vert k\vert} \cos (4k+1)\pi \theta \\ & \qquad  
+8AB \cos \pi \theta \sum\limits_k r^{\vert k\vert+\vert k-1\vert}
\cos (4k-1)\pi \theta \\
& =4\, \frac{1+r^2}{1-r^2} 
+2A^2 \, \frac{1-r^4}{1-2r^2 \cos 8\pi \theta +r^4}
+2B^2 \, \frac{(1-r^4)\cos 4\pi \theta}{1-2r^2 \cos 8\pi \theta +r^4} \\
& =\frac{4r}{1-r^2} +\frac{8AB \cos^2 \pi \theta (1-r^4)}{
1-2r^2 \cos 4\pi \theta} \\ & \qquad 
+8AB\cos^2 \pi \theta \cos 2\pi \theta\cdot \frac{2r}{1-r^2} \cdot
\frac{1-r^4}{1-2r^2 \cos 4\pi \theta +r^4} \, .
\end{split}
\end{equation*}

Hence
\begin{equation*}
\begin{split}
\| H_\theta \|^2 & \geq 4+\frac{4r}{1+r^2}+
\frac{2(A^2+B^2 \cos 4\pi \theta)(1-r^2)^2}{
1-2r^2 \cos 8\pi \theta +r^4} \\
& \qquad +8AB\cos^2 \pi \theta \bigg( 1+\frac{2r}{1+r^2} \, 
\cos 2\pi \theta \bigg) \frac{(1-r^2)^2}{1-2r^2 \cos 4\pi \theta +r^4} \, .
\end{split}
\end{equation*} 

Taking $r=\tan (\alpha /2)$ with $0<\alpha <\pi /2$, we find
\begin{equation}\label{4.5}
\begin{split}
\| H_\theta \|^2 & \geq 4+2\sin \alpha +
\frac{2(A^2+B^2 \cos 4\pi \theta )\cos^2 \alpha}{
1-\sin^2 \alpha \cos^2 4\pi \theta} \\
& \qquad +8AB \cos^2 \pi \theta (1+\sin \alpha \cos 2\pi \theta) \,
\frac{\cos^2 \alpha}{1-\sin^2 \alpha \cos^2 2\pi \theta} \, .
\end{split}
\end{equation}

Taking also
\begin{equation*}
\sin^2 \alpha =\frac{1-\vert \sin 4\pi \theta \vert}{\cos^2 4\pi \theta}
=\frac{1}{1+\vert \sin 4\pi \theta \vert} \, ,
\qquad \cos^2 \alpha =\frac{\vert \sin 4\pi \theta \vert}{
1+\vert \sin 4\pi \theta \vert} \, ,
\end{equation*}
we get
\begin{equation*}
\frac{\cos^2 \alpha}{1-\sin^2 \alpha \cos^2 4\pi \theta} 
=\frac{1}{1+\vert \sin 4\pi \theta \vert} 
\end{equation*}
and
\begin{equation*}
\frac{\cos^2 \alpha}{1-\sin^2 \alpha \cos^2 2\pi \theta} =
\frac{2\vert \cos 2\pi \theta \vert}{2\vert \cos 2\pi \theta \vert +
\vert \sin 2\pi \theta \vert} =\frac{2}{2+\vert \tan 2\pi \theta \vert} \, .
\end{equation*}

Hence \eqref{4.5} gives
\begin{equation}\label{4.6}
\begin{split}
\| H_\theta \|^2 & \geq 4+\frac{2}{\sqrt{1+\vert \sin 4\pi \theta \vert}}
+\frac{2(A^2+B^2 \cos 4\pi \theta)}{1+\vert \sin 4\pi \theta \vert} \\
& \qquad +\frac{16AB \cos^2 \pi \theta}{2+\vert \tan 2\pi \theta \vert}
\bigg( 1+\frac{\cos 2\pi \theta}{\sqrt{1+\vert \sin 4\pi \theta \vert}}
\bigg) .
\end{split}
\end{equation}

Using also the equality
\begin{equation}\label{4.7}
\max\limits_{A^2+B^2=1} 2(\alpha_0 A^2 +
\beta_0 B^2 +\gamma_0 AB )=\alpha_0+\beta_0
+\sqrt{(\alpha_0 -\beta_0)^2 +\gamma_0^2}  ,
\end{equation}
we infer from \eqref{4.6}, with
\begin{equation*}
\begin{split}
\alpha_0 & =\frac{1}{1+\vert \sin 4\pi \theta \vert} \, ,
\qquad \beta_0 =\frac{\cos 4\pi \theta}{1+\vert \sin 4\pi \theta \vert}
\\ \gamma_0  &
=\frac{8\cos^2 \pi \theta}{2+\vert \tan 2\pi \theta \vert}
\bigg( 1+\frac{\cos 2\pi \theta}{\sqrt{1+\vert \sin 4\pi \theta \vert}} 
\bigg) \, ,
\end{split}
\end{equation*}
that
\begin{equation}\label{4.8}
\| H_\theta \|^2 \geq f_2 (\theta)^2 \qquad 0\leq \theta \leq 1/2 ,
\end{equation}
with $f_2(\theta)$ given by \eqref{I11}.

If we take
\begin{equation*}
z_n =\begin{cases} \sqrt{1/10} \sin \alpha
& \mbox{ if $n=\pm 2$} \\
\sqrt{2/5} \sin \alpha & \mbox{\rm if $n=\pm 1$} \\
\cos \alpha & \mbox{\rm if $n=0$} \\
0 & \mbox{\rm if $n\neq 0,\pm 1,\pm 2$},
\end{cases}
\end{equation*}
for some $\alpha \in \R$, then $\sum\limits_n z_n^2 =1$, and so we get
\begin{equation*}
\begin{split}
\| H_\theta \|^2 & \geq 2\bigg( \frac{\sin \alpha}{\sqrt{10}} \bigg)^2
+2\bigg( \frac{2\cos 4\pi \theta \sin \alpha}{\sqrt{10}}
+\sqrt{\frac{2}{5}} \sin \alpha \bigg)^2 \\
& =2\bigg( 2\cos 2\pi \theta \sqrt{\frac{2}{5}} \sin \alpha
+\cos \alpha +\frac{\sin \alpha}{\sqrt{10}} \bigg)^2 
+\bigg( 2\cos \alpha +2\sqrt{\frac{2}{5}} \sin \alpha \bigg)^2 .
\end{split}
\end{equation*}

Since the right-hand side of the inequality above is equal to
\begin{equation*}
\begin{split}
& \frac{\sin^2 \alpha}{5} +\frac{4}{5} \, \sin^2 \alpha 
(1+\cos 4\pi \theta )^2 +2\bigg(
\cos \alpha +\frac{\sin \alpha (1+4\cos 2\pi \theta)}{\sqrt{10}}
\bigg)^2 \\
& \qquad \qquad 
+4\bigg( \cos \alpha +\sqrt{\frac{2}{5}} \sin \alpha \bigg)^2 \\
& \qquad =6\cos^2 \alpha +\frac{\sin^2 \alpha (9+16\cos^4 2\pi \theta +
(1+4\cos 2\pi \theta )^2)}{5} \\ & \qquad \qquad +\sin 2\alpha \bigg( 
\frac{2(1+4\cos 2\pi \theta )}{\sqrt{10}}+4\sqrt{\frac{2}{5}} \bigg) \\
& \qquad =6\cos^2 \alpha +2\sin^2 \alpha \bigg( 1+\frac{4}{5}
(\cos 2\pi \theta+2\cos^2 2\pi \theta+2\cos^4 2\pi \theta )\bigg) \\
& \qquad \qquad +\sin 2\alpha \bigg( \sqrt{10}+
\frac{8\cos 2\pi \theta}{\sqrt{10}} \bigg) ,  
\end{split}
\end{equation*}
we deduce from \eqref{4.7} that
\begin{equation}\label{4.9}
\| H_\theta \|^2 \geq f_3(\theta)^2 ,
\end{equation} 
where $f_3(\theta)$ is as in \eqref{I12} .

\bigskip

\bibliographystyle{amsplain}

\end{document}